\documentclass[twocolumn, switch]{article}

\usepackage{preprint}

\usepackage{amsmath, amsthm, amssymb, amsfonts, bm}

\usepackage[numbers,square]{natbib}
\usepackage[utf8]{inputenc}	
\usepackage[T1]{fontenc}	
\usepackage{xcolor}		
\usepackage[colorlinks = true,
            linkcolor = purple,
            urlcolor  = blue,
            citecolor = cyan,
            anchorcolor = black]{hyperref}	
\usepackage{booktabs} 		
\usepackage{nicefrac}		
\usepackage{microtype}		
\usepackage{lineno}		
\usepackage{float}		
\usepackage{graphicx}  
\usepackage{subfigure}
\usepackage{wrapfig}
\usepackage{multirow}

\usepackage{fancyhdr}
\usepackage{longtable}
\usepackage{parskip}
\usepackage{dcolumn}   

\usepackage{soul}      

\newcommand{\pic}[4]{\begin{center} \includegraphics[scale=#1]{#2}\label{#3}\caption{#4}\end{center}}

\newcommand{\pwisein}{\left\{ \begin{array}{ll}}
\newcommand{\pwiseout}{\end{array}\right.}

\usepackage{newfloat}
\DeclareFloatingEnvironment[name={Supplementary Figure}]{suppfigure}
\usepackage{sidecap}
\sidecaptionvpos{figure}{c}

\usepackage{titlesec}
\titlespacing\section{0pt}{12pt plus 3pt minus 3pt}{1pt plus 1pt minus 1pt}
\titlespacing\subsection{0pt}{10pt plus 3pt minus 3pt}{1pt plus 1pt minus 1pt}
\titlespacing\subsubsection{0pt}{8pt plus 3pt minus 3pt}{1pt plus 1pt minus 1pt}

\title{ThermoLIB - A Python Library for Constructing and Post-Processing Free Energy Surfaces to Extract Thermodynamic and Kinetic Properties}

\usepackage{authblk}

\author[1]{Massimo Bocus$^{*,}$}
\author[1]{Louis Vanduyfhuys$^{*,\dagger,}$}

\affil[1]{Center for Molecular Modeling, Ghent University, Technologiepark 46, Zwijnaarde 9052, Belgium}

\begin{document}

\twocolumn[ 
  \begin{@twocolumnfalse} 

\maketitle

\begin{center}
  $^{*}$ These authors contributed equally to this work\\
  $\dagger$ Email: \texttt{louis.vanduyfhuys@ugent.be}
\end{center}

\vspace{0.35cm}

\begin{abstract}
  ThermoLIB is Python/Cython library designed to be used as a post-processing tool for constructing free energy surfaces from the output of molecular simulations, transforming them between different collective variables (CVs) and extracting thermodynamic and kinetic information. ThermoLIB is available for download on GitHUB and comes with extended documentation as well as many tutorials. The implementation is based on the theory of maximum likelihood estimators and includes error bars on and full covariance matrix between all points on the free energy surface using the Fisher information matrix. The free energy surfaces can be transformed a posteriori to other collective variables, projected towards lower dimensional CV-spaces and even deprojected towards higher dimensional CV-spaces if additional information from the simulation is provided in the form of a conditional probability. Finally, one can extract usefull thermodynamic and kinetic properties such as the reaction free energy and kinetic rate constant. Error bars on the free energy surfaces are propagated throughout al these operations. We briefly illustrate the capabilities of ThermoLIB by means of some tutorials and case studies.
\end{abstract}
\keywords{Free energy surface, collective variable, tranformation, (de)projection, error estimate}
\vspace{0.35cm}
 
  \end{@twocolumnfalse} 
]

\section{Introduction}
\begin{figure*}
    \centering
    \includegraphics[width=14cm]{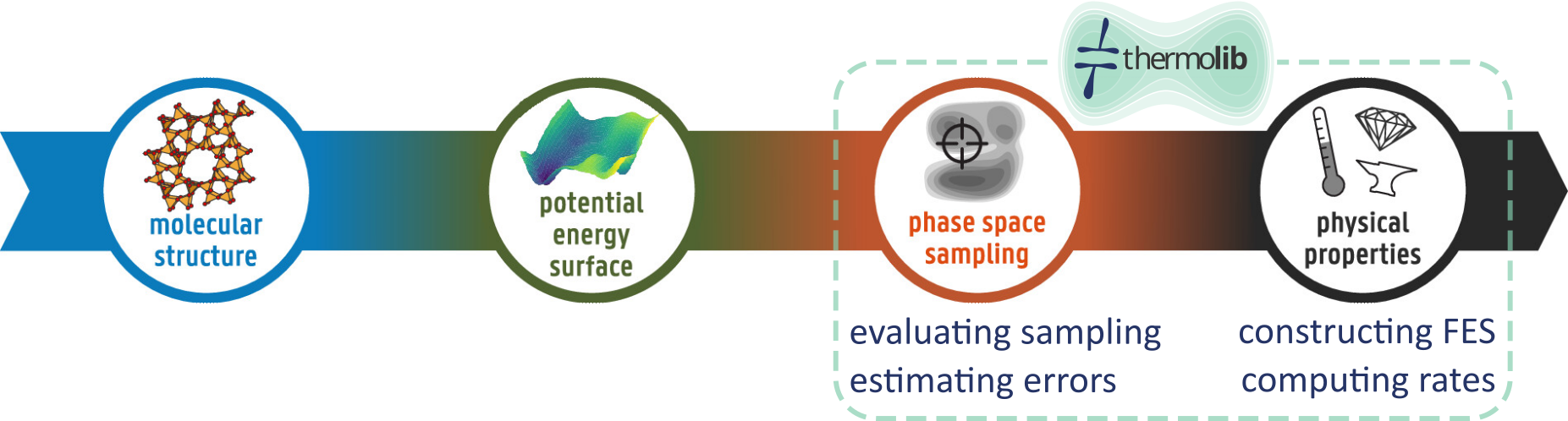}
    \caption{Typical workflow of a molecular simulation consisting of 4 steps, where ThermoLIB is a tool designed to evaluate sampling from step 3 and construct the properties from step 4. Figure adapted from Ref.\citenum{VanSpeybroeck2023}}
    \label{fig:stepsmolsim}
\end{figure*}

\begin{wrapfigure}{l}{50mm}
    \includegraphics[width=50mm]{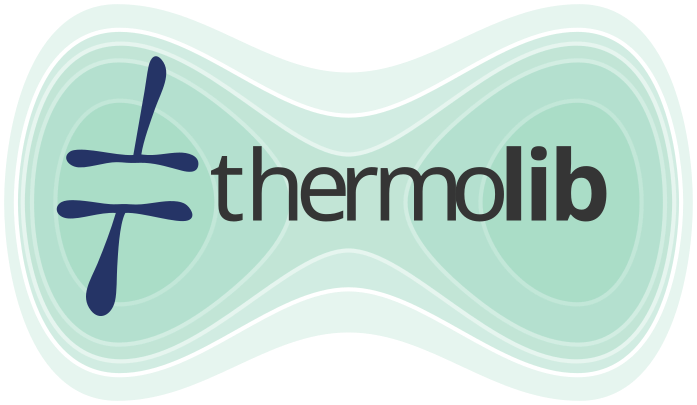}
\end{wrapfigure}
\par Computationally characterization of molecular systems by means of atomistic simulations to compute macroscopic properties has become invaluable in various research fields including chemical catalysis\cite{VanSpeybroeck2023}, material design\cite{Bosoni2024}, and drug design\cite{Decherchi2020}. The fundamental steps that need to be followed in such procedure, as highlighted by Van Speybroeck\cite{vanspeybroeck2023challenges}, are illustrated in Figure~\ref{fig:stepsmolsim}: (1) setting up a realistic molecular model for the material, (2) evaluating the potential energy surface (PES) using an adequate level of theory, (3) efficiently sampling the relevant regions of such PES, and (4) adequately extracting the macroscopic properties of interest under realistic conditions. With respect to the last step, we know from thermodynamics that knowing the free energy as function of the corresponding thermodynamic control variables (e.g., the Helmholtz free energy $F$ as function of number of particles $N$, volume $V$ and temperature $T$ for a pure $PVT$-system) allows to compute any thermodynamic property of interest. According to statistical physics, we can calculate the free energy through construction of the appropriate partition function, which is highly non-trivial due to the increasing dimensionality of the PES with system size and the associated sampling issues (step 3 in Figure~\ref{fig:stepsmolsim}).\cite{Frenkel1996} 

\par In many cases, especially when investigating physical of chemical processes that represent so-called rare events, the free energy can be conveniently partitioned in terms of a given collective variable (CV) which should quantify the progress of the process of interest. According to statistical physics, the probability that the system is in a state with CV value equal to $q$ is expressed as a boltzmann factor featuring the free energy of that state $F(q)$. Such free energy profile (FEP) $F(q)$ allows therefore to assess the thermodynamic stability of the system in terms of the CV by identifying local minima in the FEP associated with (meta)stable states. Moreover, through application of transition state theory (TST), we can extract kinetic rate constants of the process under study. In some cases, sampling the phase space (or enhancing the sampling of phase space) in terms of a single CV is not sufficient to adequately capture all relevant parts of phase space and accurately estimate reaction kinetics and $2$ or more CVs are necessary to construct a complete free energy surfaces (FES) for the process under investigation.\cite{Bocus2022} 

\par The quality of the information encoded in a free energy profile/surface largely depends on the possibility of identifying \textit{a priori} a (set of) appropriate collective variable(s) to describe the rare event, which is often a highly non-trivial task. The identification of ideal CVs is an active field of research, that has seen some significant advances with the introduction of machine learning techniques\cite{zhu2025enhanced}. Herein, we will not delve in the identification of `good' CVs, but rather focus on the construction and analysis of free energy surfaces from a predefined set of collective variables. This will include techniques to \textit{a posteriori} evaluate of the adequacy of a chosen CV and potentially improve on its sampling or even identify alternative and/or additional CVs required for adequate characterization of the process at hand. In this context, many challenges still exist that complicate a thorough and reproducible analysis of the simulation results, ranging from a robust uncertainty estimation on the predicted physical quantities to the possibility of identifying new mechanistic insights by exploring additional CVs. Implementing the tools required to address these issues is the main goal of ThermoLIB, the Python/Cython package we present in this work and is available at \url{https://github.com/molmod/ThermoLIB}. Hereafter, we outline the three major challenges we aim at addressing with ThermoLIB. 

\par Deriving a free energy profile can be as easy as a simple histogram construction from the CV samples of a single MD/MC simulation. However, the task is more challenging when multiple (unequivalently) biased simulations need to be combined. In this respect, the weighted histogram analysis method (WHAM) is one of the most widely used techniques to construct the unbiased free energy profile using input from multiple differently biased molecular simulations\cite{Kumar1992}, though alternative approaches also exist\cite{shirts2008statistically}. The WHAM equations were originally derived on the basis of the histogram method of Ferrenberg and Swendsen by means of reweighting the individual unbiased simulations with weight factors determined to minimize the statistical uncertainty of the predicted FES.\cite{Ferrenberg1989} However, to make reliable statements with respect to the stability, mechanism or kinetics of a system undergoing a process, one also needs an adequate estimation of the error bar on the FES. This is far from trivial to estimate for the WHAM-generated profile, especially in the most general case where no restrictions on the used bias potentials are imposed, which is the first challenge addressed in ThermoLIB.

\par Even though WHAM is implemented in various codes\cite{Plumed2019}, an adequate error estimation on the resulting free energy is in many cases unavailable. For example, in the original derivation, these errors were derived from the statistical uncertainty of each bin in the histogram without accounting for the uncertainty of the normalization factor of each biased simulation\cite{Kumar1992}. A notable exception in this context is g\_wham, as it includes a procedure for error estimation based on bootstrapping\cite{Hub2010}. Multiple variants were implemented, e.g. based on resampling the MD samples within each umbrella, or by resampling the full umbrella histograms themselves. However, it is not trivial to assess the underlying assumption that resampling adequately represents the sought-after errors in the original samples. In the ongoing search for reliable error estimation for WHAM-generated free energy profiles, it has been recognized that the WHAM equations can also be derived from the application of the maximum-likelihood theory\cite{Fisher1922,cowan1998l,HardleSimar2003}, as has been done by Bartels and coworkers\cite{Bartels1997}, which reveals new pathways for error estimation. In this respect, Zhu et al. have suggested a methodology for estimating error bars using a thermodynamic integration scheme of the mean restraining force, but it is only applicable for harmonic bias potentials that all have the same force constant\cite{Zhu2011}. Gallicchio et al. proposed a Bayesian analysis of the error on the WHAM solution by performing an additional Monte Carlo simulation\cite{Gallicchio2005}. This MC run samples the likelihood function in order to extract an ensemble of probability density functions (or free energy profiles) that allows to determine the error on the probability density function as the corresponding standard deviation. However, such numerical scheme can be very time consuming, especially in the case of a large number of model parameters (e.g., a large number of histogram bins). In ThermoLIB, we have implemented an analytical estimation of the error bars on the free energy profiles from the computation of the Fisher matrix on the maximum likelihood estimator. This implementation does not assume or require any specific form of the bias potentials used in the umbrella simulations and can therefore be applied to any combination of one- and/or multidimensional bias potentials. Furthermore, we also provide the necessary tools to compute the correlation time of the sampled data from the autocorrelation function, which is subsequently used in the error estimation to account for statistical inefficiencies in the sampled data.

\par
\begin{figure}[ht!]
    \centering
    \includegraphics[width=8cm]{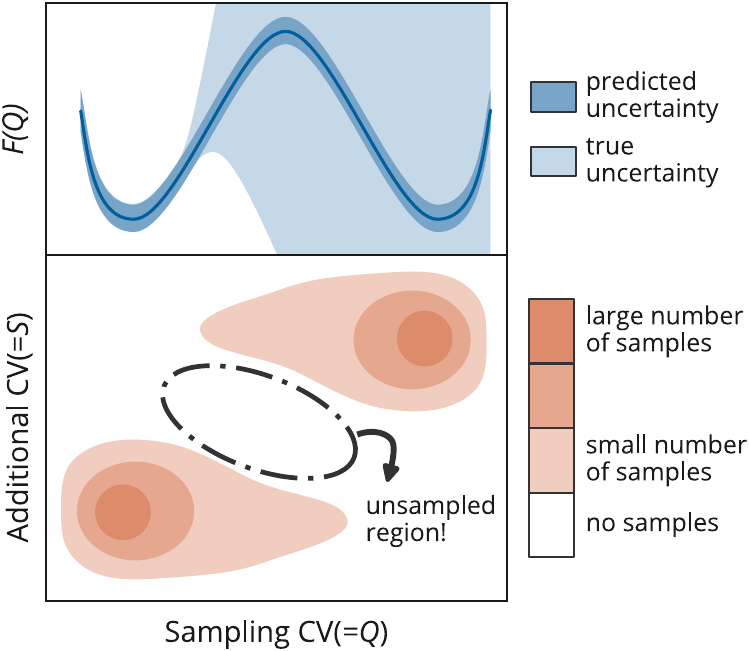}
    \caption{Illustration how a seemingly well converged 1D FEP (solid line top pane) did not result from adequate sampling. Investigating the visited states in a 2D phase space with an additional important CV (bottom pane) reveals two disjoint regions. Due to missing sampling (indicated by dot-dashed region) we have no information on the relative stability between these two parts of phase space and do not know the free energy difference between these states. In other words, the true uncertainty on the computed FEP should actually diverge to infinite once the two regions start to overlap along $Q$.}
    \label{fig:hiddencv}
\end{figure}

 The second challenge one may face upon doing molecular simulations is evaluating whether the obtained data reliably represents the relevant regions of phase space, which is an important form of sampling convergence. The error bars on the free energy surface, as discussed in the previous paragraph, is already an important tool in this respect, as it allows to detect CV regions that were undersampled. However, a more difficult problem arises when the simulation does not sample a part of phase space relevant for the process under study because of the CV selection. This could be the case even if the FEP seems to be well converged (i.e. smooth and low error bars), and was already highlighted many times before in literature\cite{dellago2002transition,barducci2011metadynamics,aho2024all} in the context of so-called hidden variables and is illustrated in Figure~\ref{fig:hiddencv}. Consider an enhanced sampling simulation performed in 1D with a single CV (denoted as $Q$) describing a process for which an additional CV can be identified based on physical/chemical intuition as potentially important to drive the process. We will denote this CV as \textit{S} and it can be recognized as a previously mentioned hidden variable. A common example of such situation is when e.g. two coordination numbers (i.e. $CN_1$ and $CN_2$) can be identified to define the process, but initial enhanced sampling is done in 1D using $Q=CN_1+CN_2$. We can then identify the orthogonal direction as the additional CV: $S=CN_1-CN_2$. In such case, we can detect whether or not there are sampling issues in terms of $S$ by doing a so-called deprojection of the original 1D FEP (in terms of $Q$) toward a 2D FES in terms of $Q$ and $S$. Regions of missing (yet crucial) sampling can be identified by the lack of samples, which can then be targeted (for example) with additional 2D umbrella simulations. In ThermoLIB, we provide the required tools to do such deprojections and identify missing regions. Furthemore, the WHAM implementation in ThermoLIB has the necessary flexibility to construct new free energy profiles/surfaces afterward using the original 1D umbrellas as well as the new additional 2D umbrella simulations.

\par Additionally, it can also be useful to simply transform a free energy profile $F(Q)$ in terms of a collective variable $Q$ to a new free energy profile in terms of a different collective variable $\tilde{Q}$. This can be useful to detect sampling issues, better separate metastable states, or fairly comparing free energy profiles obtained from different simulations as illustrated by Bailleul et al\cite{Bailleul2020}. For example, this type of transformations has been used by some of us to analyze the equivalency of FEPs obtained in function of centroid and beads from path integral molecular dynamics simulations\cite{Bocus2023,lamaire2022quantum}. ThermoLIB allows to do such transformations, both if the connection between the old and new CV is described by means of a deterministic function $\tilde{Q}=f(Q)$, as well as in the case in which it is described by a probabilistic correlation in the form of a conditional probability $p_{\tilde{Q}|Q}(\tilde{q}|q)$.

\par The third challenge lies in the extraction of reliable kinetic rate constants from molecular simulations for systems undergoing physical transitions or chemical reactions. In this respect, transition state theory (TST) is widely applied, though one must avoid the naive approach in which the min-max difference in free energy values obtained from a FEP is plugged into the 1D TST formula obtained by Eyring\cite{Hanggi1990}. Kinetic rates derived in such way will be (strongly) dependent on the used collective variable, which makes them obviously unreliable. This is a direct consequence of the fact that the free energy profile $F(q)$ inherently depends on the choice of the collective variable $Q$. As was shown by some of the present authors as well as others in literature, only a rigorous application of TST, properly accounting for the reactant state width as well as the rate of change of the CV at the transition state, will result in CV-independent rate constants\cite{Bailleul2020,Bal2020,Dietschreit2022,Dietschreit2023}. In ThermoLIB, we have implemented the required tools to reliably estimate such kinetic rate constants including error bars that are propagated from the underlying free energy profile. 

\par As such, ThermoLIB provides tools to tackle various non-trivial tasks required for the construction and manipulation of free energy surfaces, as well as an adequate estimate and interpretation of derived thermodynamic and kinetic properties including reliable error bars. In the remainder of this paper, we will focus on the three challenges above and illustrate how to apply ThermoLIB to tackle them, summarizing the methodology behind it. A full theoretic derivation of all methodologies implemented in ThermoLIB is provided in the supporting information, and an extensive manual on how to use ThermoLIB, including tutorials, is available in the online documentation at \url{https://molmod.github.io/ThermoLIB/}.

\section{Methodology}
\subsection{Free energy surfaces with adequate error estimate}
The starting point to all features in ThermoLIB is the construction of free energy surfaces in terms of an \textit{a priori} chosen set of collective variables from the output of multiple (possibly biased) molecular dynamics simulations using the Weighted Histogram Analysia Method (WHAM)\cite{Ferrenberg1889,Kumar1992}, including adequate estimates of the corresponding error bars. To allow for a clear elaboration on the theoretical background of all features implemented, it is useful to revisit the WHAM equations using the theory of maximum likelihood\cite{Bartels1997}. We refer the interested reader to the supporting information for a complete derivation. For the sake of simplicity, we will here focus on the one-dimensional case, i.e., construction of a free energy $F(q)$ as function of a single collective variable $q$, although the extension towards multiple dimensions is straightforward. The starting point is to expand the unbiased probability density $p^{(0)}(q)$ as a simple linear combination of basis functions $\phi_k(q)$:
\begin{align}
    p^{(0)}(q|\vec{a}) &= \sum_{k=1}^{K}a_k\phi_k(q) \\
    \phi_{k}(q) &= \begin{cases}
        \frac{1}{\Delta} & q\in\left[Q_{k}-\frac{\Delta}{2},Q_{k}+\frac{\Delta}{2}\right]\\
        0 & \text{elsewhere}
    \end{cases}
\end{align}
\noindent The basis functions can be chosen freely, but in this work we chose simple bin-type functions with $\Delta$ the bin width and $Q_k$ the center of bin $k$. The expansion coefficients $\vec{a}=(a_1,a_2,\ldots, a_K)$ are the sought-after parameters we will estimate using maximum likelihood. As such, $p^{(0)}(q|\vec{a})$ represents the model for the probability density in terms of $q$ given by expansion coefficients $\vec{a}$. Due to the normalization and non-overlapping nature of $\phi_k(q)$, we can interpret $a_k$ as the probability of finding the system in a state with $q$ within $\left[Q_{k}-\frac{\Delta}{2},Q_{k}+\frac{\Delta}{2}\right]$. This implies two constraints on the parameters: (1) they are positive ($\forall k: a_k \ge 0$) and (2) they are normalized ($\sum_k a_k =1$). The positivity is imposed by means of the substitution $a_k=e^{-g_k}$, giving rise to the probability distribution $p^{(0)}(q|\vec{g})$, while the normalization will be enforced using Lagrange multipliers. Next, the probability density $p^{(i)}$ of the biased simulation $i$ is expressed in terms of the unbiased probability density $p^{(0)}$ and the biasing potential $W^{(i)}$:
\begin{align}
    p^{(i)}(q|\vec{g},f_i) &= f_i p^{(0)}(q|\vec{g}) e^{-\beta W^{(i)}(q)}
\end{align}
\noindent where the additional factor $f_i$ ensures normalization of $p^{(i)}$ and represents an additional model parameter that is to be estimated. The so-called maximum likelihood estimator (MLE) $(\hat{\vec{g}},\hat{\vec{f}})$ of the model parameters is obtained by maximizing the following function $l_{\vec{N}}$: 
\begin{align}
    l_{\vec{N}}(\vec{g},\vec{f}) &= \sum_{i=1}^{M}c_i l^{(i)}_{N_i}(\vec{g},f_i) \\
    l^{(i)}_{N_i}(\vec{g},f_i) &= \frac{1}{N_i}\sum_{n=1}^{N_i}\log\left[p^{(i)}(q_n|\vec{g},f_i)\right]
\end{align}
\noindent The function $l_{\vec{N}}$ represents a weighted average of the log-likelihood functions $l^{(i)}_{N_i}$ associated with each simulation $i$. Each log-likelihood $l^{(i)}_{N_i}$ in turn represents the logarithm of the likelihood that the parameters $\vec{g}$ and $f_i$ represent the true parameters of the biased probability distribution $p^{(i)}(q|\vec{g},f_i)$ of simulation $i$, given that the samples $\{q_n\}_{n=1}^{N_i}$ were drawn from this probability distribution in the biased simulation. The subscript $N_i$ in $l^{(i)}_{N_i}$ indicates it is constructed from $N_i$ samples taken from simulation $i$. Similarly, the subscript $\vec{N}$ in $l_{\vec{N}}$ indicates this function is constructed by taking $N_1$ samples from simulations $1$, $N_2$ samples from simulations $2$, \ldots  The set of parameters $(\hat{\vec{g}},\hat{\vec{f}})$ that maximize the log-likelihood function $l_{\vec{N}}$ are denoted as the maximum likelihood estimator and it can be shown they converge towards to true parameters in the limit for infinite number of samples (see supporting information). The most sensible choice of the weights is given by $c_i=\frac{N_i}{N}$ with $N=\sum_i N_i$ (see supporting information). Finally, we proceed by maximizing $l_{\vec{N}}(\vec{g},\vec{f})$ using Lagrange multipliers to apply the normalization constraints of the unbiased and biased simulations, which results in the maximum likelihood estimators (MLE) $\hat{\vec{g}}$ and $\hat{\vec{f}}$:
\begin{align}
    e^{-\hat{g}_{k}}        & =\frac{\sum_{i=1}^{M}H_{ik}}{\sum_{i=1}^{M}N_{i}\hat{f}_{i}b_{ik}}    \label{eq:WHAM_eq1}\\
    \frac{1}{\hat{f}_{i}}   & =\sum_{k=1}^{K}b_{ik}e^{-\hat{g}_{k}}                                 \label{eq:WHAM_eq2}\\
    1                       & =\sum_{k=1}^{K}e^{-\hat{g}_{k}}                                       \label{eq:WHAM_eq3}
\end{align}
with
\begin{align}
    b_{ik} &= \frac{1}{\Delta}\int_{Q_k-\frac{\Delta}{2}}^{Q_k+\frac{\Delta}{2}} e^{-\beta W^{(i)}(q)}dq
\end{align}
\noindent and where $H_{ik}$ represents the number of $q$-samples from simulation $i$ that are located in bin $k$. These equations represent the WHAM equations, which need to be solved iteratively for $\hat{\vec{g}}$ and $\hat{\vec{f}}$. The error on this MLE can be determined from the Fisher information matrices, which express the amount of information each simulation carries on the unknown parameters $\vec{g},\vec{f}$ (as well as the Lagrange multipliers). In case there would be no normalization constraints, the Fisher information matrix for simulation $i$ would have following matrix elements:

\begin{align}
    \left[\bar{\bar{I}}^{(i)}\right]_{kl} &= -E^{(i)}\left[\frac{\partial^2}{\partial g_k\partial g_l}\log p^{(i)}(q|\vec{g})\right]
\end{align}

\noindent in which $E^{(i)}\left[f(q)\right]$ indicates the population average of $f(q)$ according to the biased probability distribution of simulation $i$. However, the normalization constraints that are enforced using Lagrange multipliers have an impact on the Fisher information matrix. First, for simulation $i$ it takes on following block matrix structure:
\begin{align}
    \bar{\bar{I}}^{(i)} &= \begin{pmatrix}
        \bar{\bar{I}}^{(i)}_{gg} & \bar{\bar{I}}^{(i)}_{gf} & \cdots \\
        \bar{\bar{I}}^{(i)}_{gf} & \bar{\bar{I}}^{(i)}_{ff} & \cdots \\
        \vdots & \vdots & \ddots 
    \end{pmatrix}
\end{align}
\noindent Herein, $\bar{\bar{I}}^{(i)}_{gg}$ encodes the amount of information present in simulation i on the parameters $\vec{g}$ and $\bar{\bar{I}}^{(i)}_{gf}$ encodes the mutual information between the $\vec{g}$ and $\vec{f}$ parameters. The dots indicate additional blocks related to information on the Lagrange multipliers. Second, the matrix elements for e.g. $\bar{\bar{I}}^{(i)}_{gg}$ are now given as:

\begin{align}
    \left[\bar{\bar{I}}^{(i)}_{gg}\right]_{kl} &= -E^{(i)}\left[\frac{\partial^2}{\partial g_k\partial g_l}\left(\log p^{(i)}(q|\vec{g})\right.\right. \nonumber\\
                                               &              + \mu\left(1-\int p^{(0)}(q|\vec{g})dq\right) \nonumber\\
                                               & \left.\left. + \lambda_i\left(1-f_i\int p^{(0)}(q|\vec{g})e^{-\beta W^{(i)}(q)}dq\right)\right)\right]
\end{align}

\noindent where the second and third line incorporate the introduction of the normalization constraints on respectively the unbiased probability distribution and the biased probability distribution of simulation $i$. Finally, we need to combine all simulations and collect the total amount of information, which can be done by computing the weighted sum of all Fisher matrices:

\begin{align}
    \bar{\bar{I}} = \sum_i \frac{N_i}{\tau_i}\bar{\bar{I}}^{(i)}
\end{align}

Herein, $\tau_i$ represents the (integrated) correlation time to account for correlation between samples. This expression basically encodes that simulations with more (uncorrelated) samples contribute more information. As a final ingredient, one can prove the so-called asymptotic normality of the MLE, meaning that in the limit for large sample sizes, the distribution of the observed MLE converges towards a normal distribution with mean given by the true parameters and covariance given by:
\begin{align}
    \bar{\bar{\sigma}} &= \bar{\bar{I}}^{-1} = \left[\sum_i \frac{N_i}{\tau_i} \bar{\bar{I}}^{(i)}\right]^{-1}
\end{align}
\noindent This covariance has the same block matrix structure as the information matrices $\bar{\bar{I}}^{(i)}$, but now we are only interested on the covariance of the parameters $\vec{g}$ of the unbiased probility distribution. However, as the inverse of the information matrix has to be computed, it is important to account for all blocks in $\bar{\bar{I}}^{(i)}$. Computation of the total information matrix $\bar{\bar{I}}$ (including accounting for the correlation between samples) and its inversion is implemented in ThermoLIB to estimate the covariance matrix on the estimated model parameters $\vec{g}$ of the unbiased probability distribution. Its diagonal elements represent the variance on each of the model parameters, from which we compute error bars on the free energy profile as a $2$-sigma interval (corresponding to $95$ \% confidence interval under the assumption of normality of the MLE). Its off-diagonal elements encode the correlation between various points on the FEP and is crucial for proper error propagation (see further below).

\subsection{Transforming and (de)projecting free energy surfaces}
In this section, we outline various methodologies on how to transform free energy profiles from one CV to another, project a 2D FES towards a 1D FEP as well as deproject a 1D FEP towards a 2D FES. First we focus on transforming a FEP $F_Y(y)$ in terms of collective variable $Y$ towards the corresponding free energy profile $F_X(x)$ in terms of another collective variable $X$, which can be done in two different ways depending on whether the relation between $Y$ and $X$ is a deterministic function or a probabilistic correlation. In the first case, the two CVs are related deterministically though the mathematical function $Y=h(X)$. By demanding that the probability $p_X(x)dx$ of finding the system in a state with $X\in \left[x,x+dx\right]$ is identical to the probability $p_Y(y)dy$ of finding the system in a state with $Y\in \left[y,y+dy\right]$ (with $y=h(x)$ and $dy=h'(x)dx$), we find the following deterministic transformation formula:
\begin{align}
    F_X(x) &= F_Y\left(f(x)\right)-k_BT\ln\left|h'(x)\right|
\end{align}
\noindent For a linear expression (with $h'(x)=\text{cte}$), the two FEPs are identical up to an irrelevant additive constant. However, for a non-linear transformation, the two FEPs can be substantially different. This transformation can be extended towards 2D:
\begin{align}
    F_X(x_1,x_2) &= F_Y\left(y_1,y_2\right) -k_BT\ln\left|J(x_1,x_2)\right|
\end{align}
\noindent with $y_1=h_1(x_1,x_2)$, $y_2=h_2(x_1,x_2)$ and associated Jacobian determinant:
\begin{align}
    J(x_1,x_2) &= \left| \frac{\partial \left[y_1,y_2\right]}{\partial \left[x_1,x_1\right]} \right| 
               = \begin{vmatrix}
                    \frac{\partial h_1}{\partial x_1} & \frac{\partial h_1}{\partial x_2} \\
                    \frac{\partial h_2}{\partial x_1} & \frac{\partial h_2}{\partial x_2}
                  \end{vmatrix}
\end{align}
\noindent A different situation arises when two CVs do not have a one-on-one connection, but are correlated through a conditional probability distribution $p_x^y(x|y)$. Application of Bayes' theorem then allows to derive the following probabilistic transformation formula:
\begin{align}
    F_x(x) &= -k_BT\ln\left[\frac{x_0}{y_0}\int_{-\infty}^{+\infty} p_x^y\left(x|y\right)e^{-\beta F_y(y)}dy\right]
\end{align}
\noindent Here, $x_0$ and $y_0$ are constants required to ensure correct dimensionality in the expression relating the probability density and free energy surface. However, as they only introduce a constant shift of free energy profiles, we will disregard them. This expression can also be extended towards a 2D transformation:
\begin{align}
    F_x(x_1,x_2) &= -k_BT\ln\left[C\iint\displaylimits_{-\infty}^{+\infty} p_x^y\left(x_1,x_2|y_1,y_2\right)e^{-\beta F_y(y_1,y_2)}dy_1dy_2\right]
\end{align}
\noindent in which $C=\frac{x_{10}x_{20}}{y_{10}y_{20}}$ is a meaningless normalization constant. Finally, we note that the deterministic transformation can be retrieved as an extreme case from the probabilitic case. In 1D this can be seen by setting $p_x^y\left(x|y\right)=\delta\left(x-f^{-1}(y)\right)$. \\

\par Aside from transforming the FES from one set of CVs to another, one can also project a high-dimensional FES towards a lower dimensional FES. To illustrate this, we consider the situation of projecting a 2D FES in terms of the CVs $(X,Y)$ to a 1D FEP in terms of a single CV $Q$. The projection formula can be obtained from the conditional probability $p_Q^{XY}(q|x,y)$ between $Q$ and $(X,Y)$:
\begin{align}
    p_Q(q) &= \iint p_Q^{XY}(q|x,y)p_{XY}(x,y)dxdy
\end{align}
\noindent Using $p_Q=\frac{\exp(-\beta F_Q)}{Z}$ and $p_{XY}=\frac{\exp(-\beta F_{XY})}{Z}$, we arrive at the desired projection formula:
\begin{align}
    F_Q(q) &= -k_BT\ln\left[C\iint p_Q^{XY}(q|x,y)e^{-\beta F_{XY}(x,y)}dxdy\right]
\end{align}
\noindent with $C=\frac{q_0}{x_0y_0}$ in which the constants $q_0$, $x_0$ and $y_0$ are again meaningless dimensionality constants that we will disregard. We can use a similar reasoning to obtain the following deprojection formula:
\begin{align}
    p_{XY}(x,y) &= \iint p_{XY}^Q(x,y|q)p_Q(q) dq \\
    F_{XY}(x,y) &= -k_BT\ln\left[C\iint p_{XY}^Q(x,y|q) e^{-\beta F_Q(q)}dq\right]
\end{align}
\noindent with $C=\frac{x_0y_0}{q_0}$. As such, the main ingredient required to deproject a 1D FEP in terms of a single CV $q$ towards a 2D FES in terms of two CVs $(x,y)$, is the conditional probability $p_{XY}^Q(x,y|q)$ expressing the probability to find a system in a state with $(X,Y)=(x,y)$ on the condition that the system already was in a state with $Q=q$. This conditional probability can be derived from a molecular simulation, even if it was biased along CV $Q$ (as shown in the supporting information).

\subsection{Extracting kinetic rate constants}
In order to extract kinetic rate constants from molecular dynamics simulations, we start from the Bennett-Chandler expression\cite{Bennett75,Chandler78} given by:
\begin{align}
    k(t) &= \frac{\left\langle\dot{q}(0)\theta\left(q(t)-q^*\right)\delta\left(q(0)-q^*\right)\right\rangle}{\left\langle\theta\left(q^*-q\right)\right\rangle}
\end{align}
\noindent At this point, $q^*$ simply represents the CV value for the border between reactant and product state. As long as this value is situated in a region of high free energy (i.e.~with a probability that is negligible w.r.t.~that of the reactant and product states), the resulting rate constant is insensitive towards its specific choice. A typical choice is the local maximum of the FEP in between the reactant and product states, as this will later also corresponds to the point-of-no-return crucial within transition state theory. As a first step, we rewrite this expression to identify contributions to the rate factor that can be extracted directly from the FEP $F_Q$\cite{VanSpeybroeck2023,Bucko2017}:
\begin{align}
    k(t) &= A(t)\frac{e^{-\beta F_Q(q^*)}}{\int_{-\infty}^{q^*} e^{-\beta F_Q(q)}dq} \label{eqn:meth_rate_k_cb2}
\end{align}
\noindent Where the prefactor $A(t)$ is given by
\begin{align}
    A(t) &= \frac{\left\langle\dot{q}(0)\theta\left(q(t)-q^*\right)\delta\left(q(0)-q^*\right)\right\rangle}{\left\langle\delta\left(q(0)-q^*\right)\right\rangle} \\
         &= \left\langle\dot{q}(0)\theta\left(q(t)-q^*\right)\right\rangle_{q(0)=q^*}
\end{align}
\noindent and encodes the rate of change of $q$ at time $t=0$ assuming it started from $q^*$ at $t=0$ and ended up in the product state at time $t$. Next, we apply transition state theory, in which we assume the dividing surface $q=q^*$ at the local maximum of the $F_Q$ profile is interpreted as the point of no return, i.e.~once the system crosses this point from the reactant state towards the product state it will always end up in the product state. In other words, recrossing back towards the reactant state is neglected. As a result of this assumption, one can take the limit of $k(t)$ and $A(t)$ for $t\rightarrow 0^+$, which results in:
\begin{align}
    k^\text{TST} &= \lim_{t\rightarrow 0^+} k(t) = A^\text{TST}\frac{e^{-\beta F_Q(q^*)}}{\int_{-\infty}^{q^*} e^{-\beta F_Q(q)}dq} \label{eqn:rate_TST_final}\\
    A^\text{TST} &= \lim_{t\rightarrow 0^+} A(t) = \frac{1}{2}\left\langle\left|\dot{q}\right|\right\rangle_{q=q^*}
\end{align}
As a final step, one can simplify the expression of the prefactor $A^\text{TST}$ by analytically performing the integration over momenta in the ensemble average\cite{Hanggi1990}:
\begin{align}
    A^\text{TST} &= \sqrt{\frac{k_BT}{2\pi}}\left\langle\left|\vec{\nabla}_xQ\right|\right\rangle_{q=q^*} 
    \label{eqn:rate_TST_prefactor}
\end{align}
\noindent in which $\vec{\nabla}_xQ$ represent the partial derivatives of the CV towards mass-weighted cartesian coordinates. There are two additional important remarks to be made about the kinetic rate constant and its expression. First, in order to estimate error bars on the kinetic rate constant, we need error bars on the prefactor (we already discussed how to extract error bars on the free energy profile). As this prefactor is a simple ensemble average, which we will approximate by its N-sample estimator, we can apply block averaging to compute error bars taking into account possible correlations between samples\cite{AllenTildesly1989}. Second, as illustrated before in literature, if one performs enhanced sampling simulations using two different CVs, the resulting free energy profiles will likely be very different. However, the rate constant computed using Eq.~\ref{eqn:rate_TST_final} will be identical (within its sampling error bar)\cite{Bailleul2020}, provided that both CVs are `good' in describing the rare event. Therefore, to quantify kinetics in terms of a free energy barrier, we cannot use the simple min-max free energy difference approach, but it is recommended to use the phenomenological free energy barrier, which basically just reformulates the rate constants in energy units\cite{Bucko2017,VanSpeybroeck2023}:

\begin{align}
    \Delta F_\text{phen} &= -k_BT\ln\left(\beta h k^\text{TST}\right)
\end{align}

\subsection{Error propagation\label{Sec:error_propagation}}
Consider a property $Y$ that can be expressed as a function(al) of the FEP $F(q)$ as well as additional quantities $\{a_i\}_{i=1}^k$. To adequately estimate error bars on $Y$, we implemented a simple Monte Carlo scheme that takes random samples from the error distributions of the FEP and $\{a_i\}_{i=1}^k$, use them to generate samples of $Y$ and finally extract the mean and standard deviation to define the error distribution of $Y$. From the derivation of the error on the WHAM equations as an MLE, we know that the error distribution on the FEP is given by a multivariate normal distribution with a mean given by the solution of the WHAM equations and a covariance matrix estimated from the inverse of the Fisher information matrix. Hence, assuming we also know the error distribution on the additional properties $\{a_i\}_{i=1}^k$, we can estimate $95$\% confidence intervals on the estimate of $Y$. To illustrate this procedure, consider the reaction rate $k^\text{TST}$ in Eq.~\ref{eqn:rate_TST_final} expressed as a functional of the FEP and proportional to the prefactor defined in Eq.~\ref{eqn:rate_TST_prefactor}. As such, we need to generate sets of samples $\{F_{(n)}(q),A^\text{TST}_{(n)}\}_{n=1}^N$, with $F_{(n)}(q)$ a sample of the entire free energy profile (taken from its multivariate normal error distribution) and $A^\text{TST}_{(n)}$ a sample of the prefactor (whose error we assume to be normally distributed with a standard deviation estimated by applying block averaging on the ensemble average in Eq.~\ref{eqn:rate_TST_prefactor}). For each generated sample $(F_{(n)}(q),A^\text{TST}_{(n)})$, we compute the corresponding sample $k_{(n)}^\text{TST}$ for the reaction rate according to Eq.~\ref{eqn:rate_TST_final}. Finally, we compute the mean and standard deviation from $\{k_{(n)}^\text{TST}\}_{n=1}^N$ and construct a $95$\% confidence interval after adequately choosing the error distribution on $k^\text{TST}$ (usually a log-normal distribution, given that the error on the FEP dominates the error on the prefactor).

\section{Tutorials}
In this section we will provide various tutorials to illustrate the usage of ThermoLIB. We will provide code snippets which include the fundamental ThermoLIB commands. Some commands (such as function imports) will be omitted for clarity. The full code can be found in corresponding notebooks in the \verb|tutorials| folder of the ThermoLIB repository. In this manuscript, we only include two tutorials, while more can be found in the repository.

\subsection{Tutorial 1 - 1D FEP from umbrella sampling}
\begin{figure}[ht!]
    \centering
    \includegraphics[width=8cm]{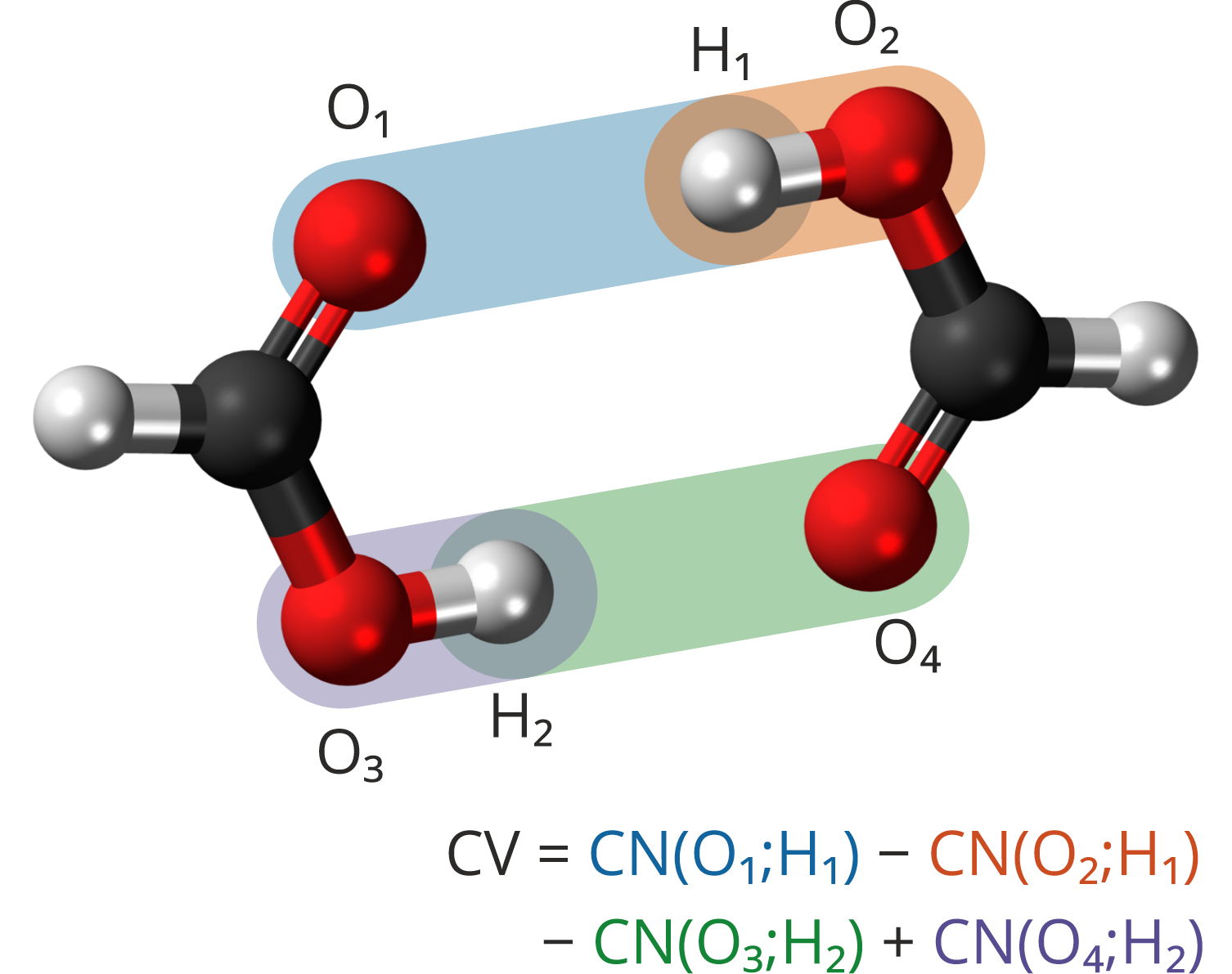}
    \caption{Illustration of a formic acid dimer and the coordination numbers used in the definition of the collective variable\label{fig:formic_acid_dimer_CV}}
\end{figure}

This tutorial demonstrates the use of ThermoLIB to construct a one-dimensional free energy profile from umbrella sampling simulations including error estimation. The example data corresponds to the proton transfer in a formic acid dimer (see figure \ref{fig:formic_acid_dimer_CV}). The collective variable is defined as a linear combination of coordination numbers (CNs) and the simulations were conducted using the same setup as in Ref.~\citenum{lamaire2022quantum}.

\noindent The code snippet below illustrates how to read the umbrella sampling trajectories and bias potentials using the \verb|read_wham_input| function. This function requires a \verb|TrajectoryReader| instance to read the CV samples from the trajectory files. In this case, we use the \verb|ColVarReader| class, which is able to read COLVAR files generated by e.g. PLUMED. The code snippet also shows how to specify the units of the CV and bias potential parameters.

\begin{verbatim}
    cvreader = ColVarReader(
    	[0], units=['au']
    )
    temp, biasses, trajs = read_wham_input(
        'wham_input.txt', cvreader, 
        'colvars/COLVAR_%s.dat', 
        bias_potential='Parabola1D', 
        q01_unit='au', 
        kappa1_unit='kjmol',
    )
\end{verbatim}

The function \verb|read_wham_input| requires the following inputs:

\begin{itemize}
    \item The metadata file (\verb|wham_input.txt| in this example) which defines the bias potentials used in each umbrella and, optionally, the simulation temperature. In case of 1D umbrella sampling, each line of this file is of the form `\verb|label kappa1 q01|', where \verb|label| is an identifier for the umbrella, \verb|kappa1| the force constant of the bias potential and \verb|q01| the center of the bias potential. Optionally, a line defining the temperature can be included in the form `\verb|temp T|', where \verb|T| is the temperature in Kelvin. If no temperature is defined, a default value of 300 K is assumed.
    \item The variable \verb|trajectory_path_template| (\verb|colvars/COLVAR_%s.dat| in the example above). This should contain a placeholder \verb|%s| that, upon substitution with the umbrella label used in the metadata file, represents the path to the trajectory file of that umbrella (i.e. a COLVAR or XYZ file).
    \item A \verb|TrajectoryReader| instance (\verb|ColVarReader| in the example above) which allows to read the trajectory files for each umbrella to extract the corresponding CV samples.
    \item The bias potential type used in the simulations (\verb|Parabola1D| in the example above).
\end{itemize}

\begin{figure*}[!h]
    \centering
    \includegraphics[width=18cm]{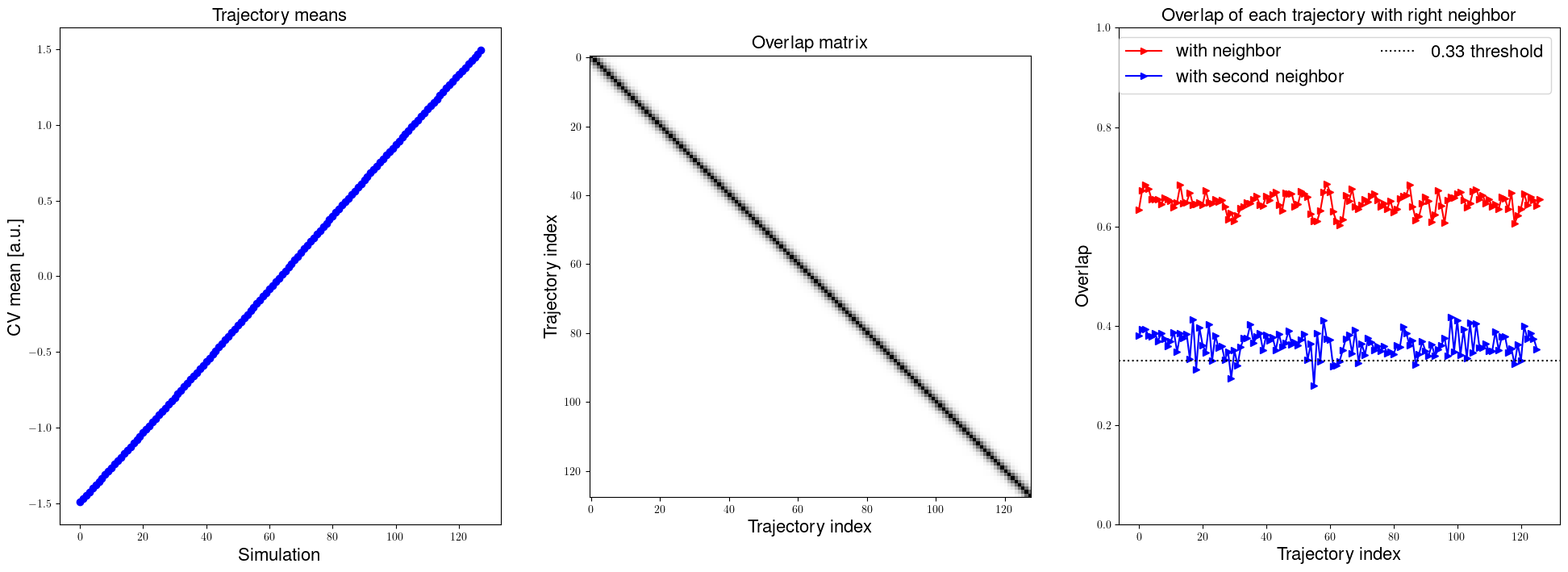}
    \caption{Trajectory analysis of the various umbrella simulations. (a) Mean of CV for each umbrella simualtion allowing to interpret which part of phase space each simulation covers (b) Overlap matrix between the histograms of each pair of umbrella simulations (c) Overlap of each umbrella simulation with its right neighbor (in red) or second right neighbor (in blue). The overlap threshold of $0.33$ is arbitrary, but represents (in general) a good rule of thumb.}
    \label{fig:tut1_histogram_overlap}
\end{figure*}

Internally, ThermoLIB consistently works in atomic units. In this case, the choice is irrelevant as CNs are unit less and so is the CV. On the other hand, should the CV be---for example---a distance between two atoms, one would need to specify \verb*|units=['angstrom']|; ThermoLIB would in that case expect a CV in \AA{}ngstrom, which would be internally converted in Bohr for all subsequent analysis. As a first validation of the simulation results, we assess the histogram overlap of neighboring umbrellas using the metric introduced by Borgmans et al. (see Eq.~5 in Ref.~\citenum{Borgmans2024}). This is done using the code given below and produces Fig.~\ref{fig:tut1_histogram_overlap}.

\begin{verbatim}
    plot_histograms_overlap_1d(trajs)
\end{verbatim}

\begin{figure}
    \centering
    \includegraphics[width=8.5cm]{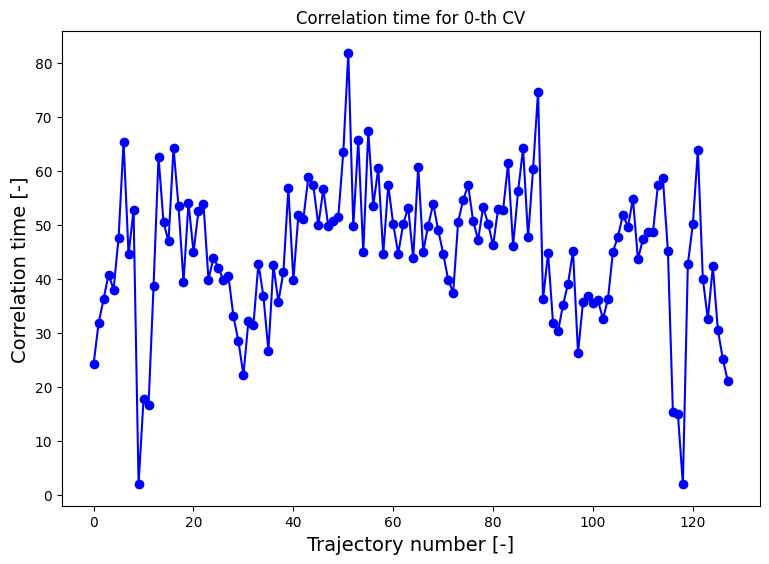}
    \caption{Autocorrelation time of the CV samples for each umbrella simulation.}
    \label{fig:tut1_corrtimes}
\end{figure}

Before we continue to construct the free energy profile using WHAM, we estimate the autocorrelation time of the CV samples. This is implemented in the \verb|decorrelate| function, which takes all umbrella sampling trajectories as argument and returns the estimated autocorrelation time for each trajectory (as well as plotting it, if requested). The code below shows how to do this and the generated plot is shown in Fig.~\ref{fig:tut1_corrtimes}.

\begin{verbatim}
    corrtimes = decorrelate(trajs, plot=True)
\end{verbatim}

\begin{figure}[!ht]
    \centering
    \includegraphics[width=8.5cm]{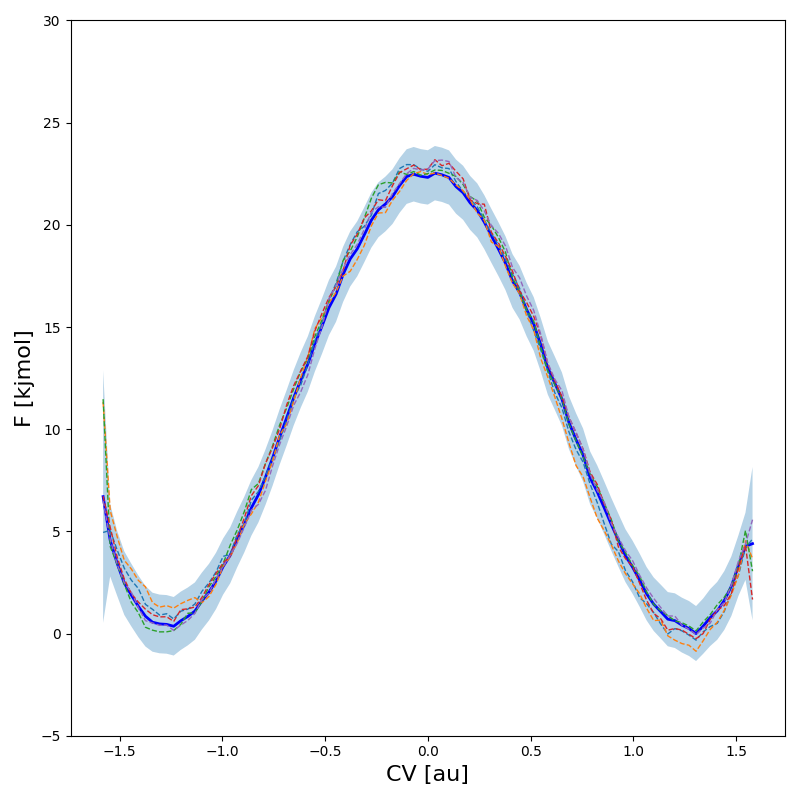}
    \caption{Free energy profile of the proton transfer reaction in the formic acid dimer. The blue line is the mean free energy profile, while the shaded area represents the 95\% confidence interval. The dashed lines are random samples of the free energy profile within its error bars.}
    \label{fig:tut1_fep}
\end{figure}

Finally, we can construct a histogram using WHAM by means of the \verb|Histogram1D.from_wham| function. This function requires the histogram bins, umbrella sampling trajectories, bias potentials and temperature as mandatory input. In this example, we also specify to estimate the error using the \verb|mle_f_cov| method, with correlation times given by \verb|corrtimes| as computed previously. The \verb|mle_f_cov| method is the advised default method to compute not only the standard deviation on the probability at each CV point, but also the covariance between the probabilities corresponding to two different CV points. To allow for reliable generation of random samples of the full free energy profile, it is required to also compute the full covariance matrix. Afterwards, we convert the histogram to a free energy profile using the \verb|BaseFreeEnergyProfile.from_histogram| routine, which also propagates the error bars and covariance. We can make a plot of the free energy profile and include random samples of the entire profile within its error bar. The code below illustrates how to do this and the generated plot is shown in Fig.~\ref{fig:tut1_fep}.

\begin{verbatim}
    bins = np.linspace(-1.7, 1.7, 100)
    hist = Histogram1D.from_wham(
        bins, trajectories, biasses, 
        temp, error_estimate='mle_f_cov', 
        corrtimes=corrtimes
    )
    fep = BaseFreeEnergyProfile.from_histogram(
        hist, temp=temp,
        cv_output_unit='au'
    )
    fep.set_ref(ref='min')
    fep.plot(
        obss=['mean']+['sample',]*5, 
        linestyles=['-']+['--',]*5,
        linewidths=[2]+[1,]*5,
        colors=['b']+[None,]*5,
        flims=[-5,30], show_legend=False,
    )
\end{verbatim}

\begin{figure}[!ht]
    \centering
    \includegraphics[width=8.5cm]{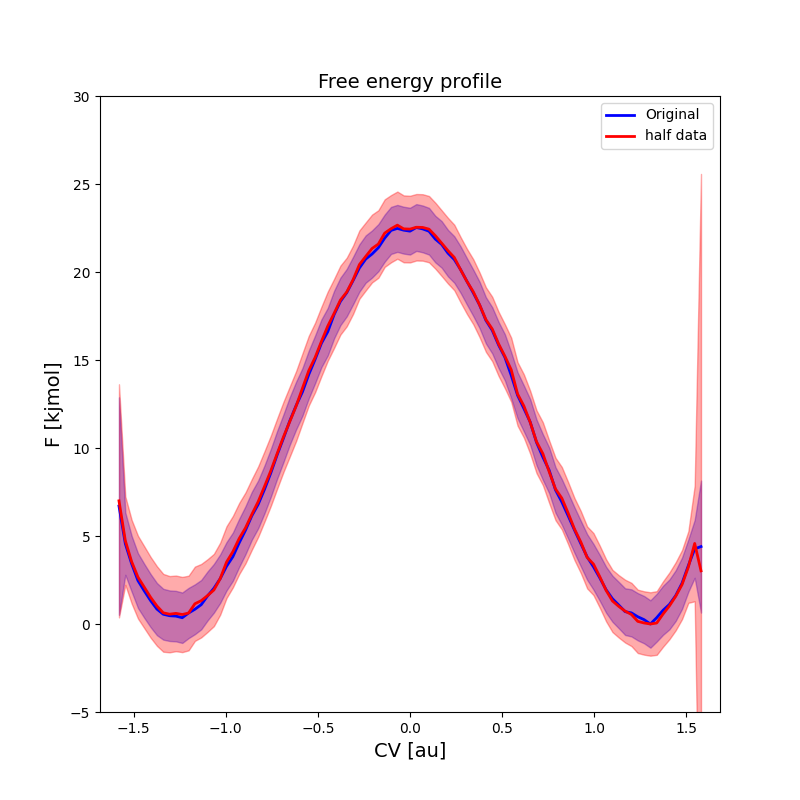}
    \caption{Comparison of the free energy profile using all umbrella sampling simulations (blue) and using only every other simulation (red). The shaded area represents the 95\% confidence interval.}
    \label{fig:tut1_fep2}
\end{figure}

Finally, to assess the rule of thumb for the overlap threshold of $0.33$, we reconsider Fig.~\ref{fig:tut1_histogram_overlap}. We observe that the overlap between second nearest neighbors (blue line) still gives rise to an overlap just above the threshold of $0.33$. Therefore, we recompute the free energy profiles using only every other trajectory from the umbrella sampling simulations. The code below shows how to do this and the generated plot is shown in Fig.~\ref{fig:tut1_fep2}.

\begin{verbatim}
    hist2 = Histogram1D.from_wham(bins, 
        trajectories[::2], biasses[::2], 
        temp, error_estimate='mle_f_cov', 
        corrtimes=corrtimes[::2]
    )
    fep2 = BaseFreeEnergyProfile.from_histogram(
        hist2, temp=temp,
        cv_output_unit='angstrom'
    )
    fep2.set_ref(ref='min')
    plot_profiles(
        [fep, fep2], 
        labels=['Original','half data'], 
        colors=['b','r'], flims=[-5,30]
    )
\end{verbatim}

As we can see in Fig.~\ref{fig:tut1_fep2}, the profile still looks almost identical (maybe slightly less smooth) as the original. We do observe a a slightly increased error bar, which is logical as it has less data and hence more uncertainty. From this, we conclude that as a rule of thumb we can state the overlap between histograms of least $0.33$ or higher seems to be appropriate for WHAM analysis.

\subsection{Tutorial 2 - Rate constant from umbrella sampling}

\begin{figure}[!ht]
    \centering
    \includegraphics[width=8.5cm]{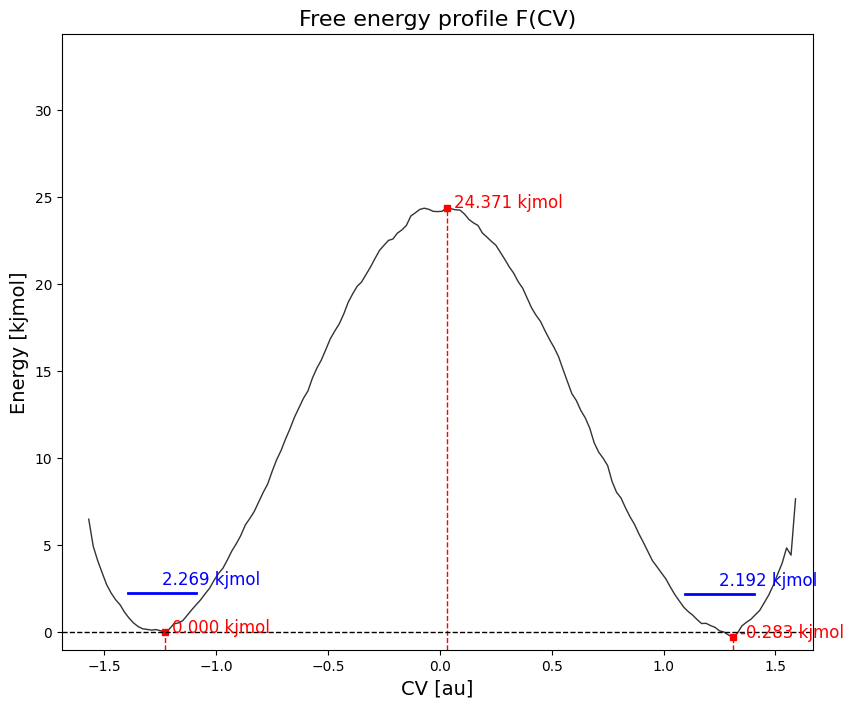}
    \caption{Free energy profile with reactant and product macrostates indicated by blue lines and transition state microstate indicated by a red dot. The corresponding free energies of these states are also indicated.}
    \label{fig:tut1_fep_withstates}
\end{figure}

After having obtained a FEP from an US simulation, we might be interested in obtaining a CV-independent rate constant for the rare event under investigation. To this end, we must compute two contributions: (1) the thermodynamic contribution given by the ratio in Equation \ref{eqn:rate_TST_final}, and (2) the rate prefactor. To achieve this, we must first identify the reactant state, product state and transition state on the FEP, which can be done using the \verb|process_states| function. Its argument `\verb|lims=[a,b,c,d]|' is used to locate the reactant microstate as local minimum of region `\verb|[a,b]|', transition microstate as local maximum of region `\verb|[b,c]|' and product microstate as local minimum of region `\verb|[c,d]|'. Additionally, the regions `\verb|[a,t]|' and `\verb|[t,d]|', with `\verb|t|' the transition microstate, are identified as reactant and product macrostates, respectively. In this context, microstates are single CV points (local minima or maxima) while macrostates are regions in CV space. In the example of this tutorial, we specify that the reactant microstate should be below $-0.2$ au, the product state above $0.2$ au and the transition state in between. The code snippet below illustrates how to do this:

\begin{verbatim}
	fep.process_states(
	    lims=[-np.inf,-0.2,0.2,np.inf]
	)
\end{verbatim}

After running this routine and replotting the free energy profile, we obtain Fig.~\ref{fig:tut1_fep_withstates}. This plot indicates three microstates (individual CV points) as red dots and two macrostates (CV regions) as blue lines. The corresponding free energy of these states are also indicated in the plot. In case of a macrostate, the free energy is computed from the integral of the boltzmanfactor over the entire macrostate region. This gives all information required to compute the thermodynamic contribution to the rate constant using Equation \ref{eqn:rate_TST_final}:

\begin{align}
    \frac{e^{-\beta F(q^*)}}{\int_{-\infty}^{q^*}e^{-\beta F(q)}dq} &= \frac{e^{-\beta\cdot(24.371\ \text{kJ/mol})}}{e^{-\beta\cdot(2.269\ \text{kJ/mol})}}
\end{align}

Second thing we must do to obtain a reaction rate, is compute an estimate of the rate prefactor (Equation \ref{eqn:rate_TST_prefactor}). In other words, we must compute the partial derivatives of the collective variable with respect to the  mass-weighted atomic coordinates. ThermoLIB offers an extensive set of implemented CVs together with their derivatives. For our formic acid dimer, we can define the linear combination of coordination numbers as follows:

\begin{verbatim}
	r0 = 1.4 * angstrom
	cn1 = CoordinationNumber([[0,1]], r0=r0)
	cn2 = CoordinationNumber([[1,2]], r0=r0)
	cn3 = CoordinationNumber([[3,4]], r0=r0)
	cn4 = CoordinationNumber([[4,5]], r0=r0)
	cv = LinearCombination(
	    [cn1, cn2, cn3, cn4], 
	    [1., -1., -1., 1.]
	)
\end{verbatim}

We can now define a \verb*|RateFactorEquilibrium| instance, which will be used to compute the overall rate. Aside from the CV definition, we must provide a small interval around the TS CV value $q^*$ to extract trajectory frames located (approximately) atop the TS.

\begin{verbatim}
	cv_ts_delta = 0.05
	cv_ts = fep.ts.get_cv()
	cv_ts_lims = [
	    cv_ts - cv_ts_delta, 
	    cv_ts + cv_ts_delta
	]
	rate = RateFactorEquilibrium(
	    cv, cv_ts_lims, temp, 
	    CV_unit='au'
	)
\end{verbatim}

We must now identify a trajectory containing a (ideally large) number of structures located within \verb*|cv_ts_lims|. The trajectory is fed to \verb*|RateFactorEquilibrium| to extract the values of the rate prefactor. Assuming we want to extract the TS samples from the trajectory of umbrella labeled as `\verb|u64|', we use the code below:

\begin{figure*}[!ht]
	\centering
	\includegraphics[width=15cm]{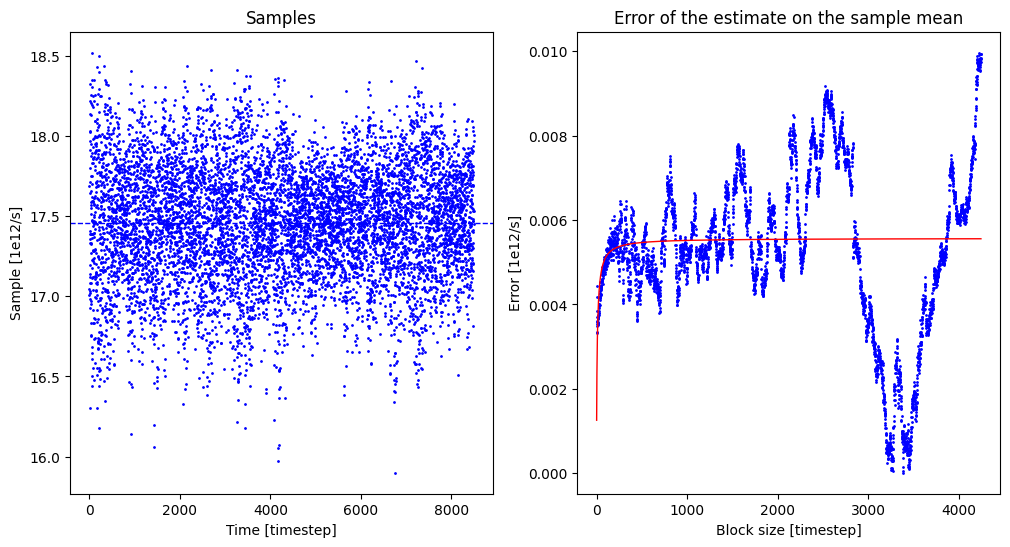}
	\caption{Values of the rate prefactor sampled from an umbrella sampling simulation, together with the block averaging results for uncertainty estimation.}
	\label{fig:rate_blav}
\end{figure*}

\begin{verbatim}
	rate.process_trajectory(
	    'trajectories/u64.xyz',
	    finish=True,
	    momenta='analytical'
	)
\end{verbatim}

Note that more than a single trajectory can be used, which is the reason for which the rate has a \verb*|finish| attribute signaling that the samples collection is terminated and the resulting averages can be computed. For multiple trajectories, one would set \verb*|finish=False| in the \verb*|process_trajectory| function and manually call \verb*|rate.finish()| once all trajectories have been processed. The error estimate on the rate prefactor can now be obtained from block averaging, where we use the empirical expression $BE = TE \sqrt{B / (B + \tau - 1)}$ to fit the computed block error values ($BE$) as a function of the block size ($B$) and obtain an estimate of the true error ($TE$) and correlation size ($\tau$, though this value is not further used). The modular nature of ThermoLIB makes the implementation of alternative expressions to obtain an estimate of $TE$ conveniently simple. The block averaging method can be called using:

\begin{verbatim}
	A, A_dist = rate.result_blav(
	    plot=True, 
	    fitrange=[1, 300]
	)
\end{verbatim}

This returns the plot shown in Figure \ref{fig:rate_blav}, where both the samples of the rate prefactor as well as the block averaging results are reported. Finally, we can obtain the overall rate results by combining the rate prefactor with the pre-computed FEP using the \verb*|compute_rate| function:

\begin{verbatim}
	rate_results = rate.compute_rate(
	    fep, verbose=True,
	)
\end{verbatim} 

The \verb*|verbose| option prints the rate results, which include the estimates of the forward and backward rate constants together with their respective phenomenological free energies. Note that the former are log-normally distributed (hence the average value does is not located at the center of the uncertainty boundaries) while the latter are normally distributed. In this example, we obtained:

\begin{verbatim}
	k_F  = 7.923 <= 23.151 <= 53.790 1e8/s
	dF_F = 17.980 +- 2.174 kjmol
	k_B  = 10.913 <= 23.152 <= 43.569 1e8/s
	dF_B = 17.856 +- 1.571 kjmol
\end{verbatim}

Reassuringly, both the rates and phenomenological free energies are identical for the forward and backward process within the error estimates.

\section{Case studies}
The alpha version of ThermoLIB has already been successfully employed in several case studies, including homo- and heterogeneous catalysis\cite{Bailleul2020,Bocus2022,bocus2022insights,Bocus2023,bocus2025operando,bocus2025confined} as well as polymer chemistry \cite{van2020cation}. Hereafter, we summarize two case studies showcasing how ThermoLIB can be very effective in highlighting and mitigating sampling issues and in exploring the mechanistic features of reactions conducted in complex environments.

\subsection{Case study 1 -  ethene chemisorption in zeolites}
\begin{figure*}
	\centering
	\includegraphics[width=13cm]{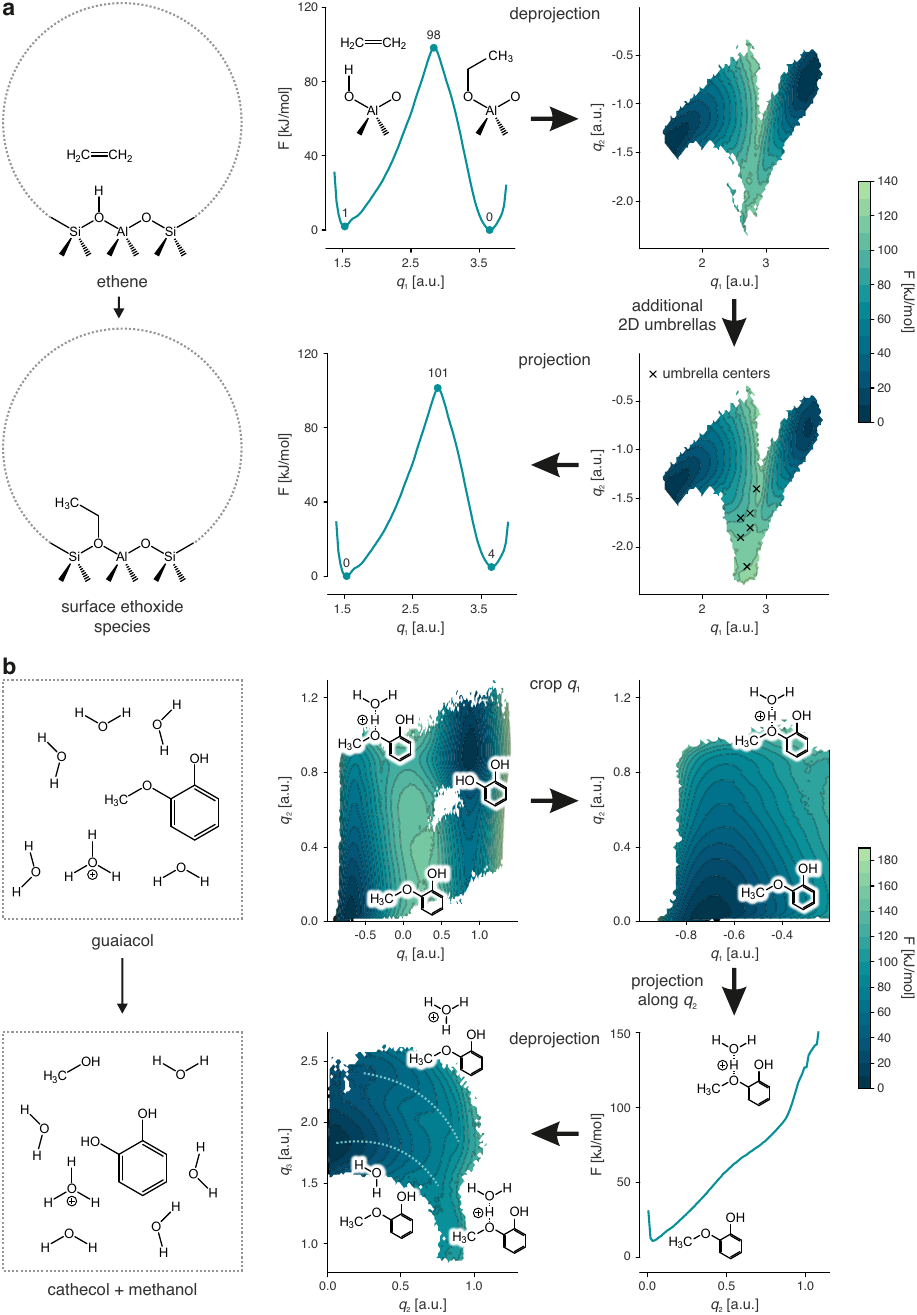}
	\caption{a. The formation of a surface-bound ethoxide species in H-ZSM-5 from physisorbed ethene. The original 1D FEP is first deprojected in a 2D FES to highlight the insufficiently sampled regions. Additional 2D umbrellas are added around the transition state to obtain a new, more complete 2D FES. The latter is then reprojected along $q_1$ to compute kinetic and thermodynamic properties. b. The concerted \textit{O}-activated S\textsubscript{N}2 demethylation of guaiacol in hot-pressurized water. An original 2-dimensional FES is first cropped to focus the investigation of the path leading to the transition state. The cropped FES is then projected along $q_2$ and deprojected along a new CV $q_3$. In this new space it is clear that two reactive channels lead to the transition state, one in which guaiacol interacts with a neutral water molecule and one in which guaiacol interacts with an hydronium ion (H\textsubscript{3}O\textsuperscript{+}).}
	\label{fig:case_studies}
\end{figure*}

In a previous work, we have used enhanced sampling ab initio molecular dynamics to investigate the ethylation of benzene with ethene and ethanol in the H-ZSM-5 zeolite.\cite{Bocus2022} To fully understand the reaction mechanism, it is essential to determine the adsorption state of ethene in the material. Indeed, the molecule can exist in a physisorbed state or it can react with the Br\o{}nsted acid sites (BASs) in the zeolite to form a surface-bound ethoxide species, as shown in Figure \ref{fig:case_studies}a. The reaction involves a concerted proton transfer to ethene together with the formation of the C-O bond. For this reason, a relatively complex linear combination of coordination numbers is required to include both processes and the chemical equivalency of the atoms involved in the reaction. In the study, the authors resorted to the following CV: $q_1=\mathrm{CN(C;O)}+0.5(\mathrm{CN(C;H)}-\mathrm{CN(O;H)})$. They then deprojected the original FEP in a 2-dimensional FES along the orthogonal CV $q_2=\mathrm{CN(C;O)}-0.5(\mathrm{CN(C;H)}-\mathrm{CN(O;H)})$. This highlighted that the process is inherently 2-dimensional and the transition state region was inadequately sampled by the original 1-dimensional umbrellas.

For this reason, they added several 2-dimensional umbrellas (i.e., biased along both $q_1$ and $q_2$) and combined them with the original data to enrich the sampling in the TS region. The newly obtained FES can then be reprojected along $q_1$ to obtain a more reliable FEP that can be used for further kinetic and thermodynamic analysis.

\subsection{Case study 2 -  guaiacol demethylation}
Guaiacol is a model compound for lignin-derived molecules in biomass conversion processes. En route towards commodity chemicals, guaiacol can be demethoxylated to cathecol using hot-pressurized water with an homogeneous acid catalyst or in Br\o{}nsted acidic zeolites. A recent study\cite{bocus2025confined} reports a complete investigation of the demethoxylation mechanism in both environments using enhanced sampling ab initio MD. The reactivity is dominated by a 1-step O-activated S\textsubscript{N}2 reaction. The authors used 2-dimensional umbrella sampling to investigate the mechanism, following both the S\textsubscript{N}2 reaction and the protonation state of the methoxy oxygen atom by means of coordination numbers and linear combinations thereof (Figure \ref{fig:case_studies}b).

While this allowed to discover that no metastable protonated intermediate is present at the onset of the S\textsubscript{N}2 step, the authors wanted to understand at which point the proton is brought in proximity of the methoxy oxygen by the water, in the form of a hydronium ion (H\textsubscript{3}O\textsuperscript{+}). To this end, the original FES was first cropped to remove the product region, focusing exclusively on the path leading to the transition state. The cropped FES was then projected along $q_2$, which follows the protonation degree of the guaiacol methoxy oxygen, and further deprojected along a third CV $q_3$. The latter monitors the coordination number between the water oxygen closest to the methoxy oxygen and all the protons in the system. The newly obtained FES clearly highlights how two reactive channels exist that lead to the transition state, one in which the hydronium ion is from the beginning in contact with guaiacol and one in which guaiacol interacts with neutral water molecules up to the onset of the transition state. 

By repeating this analysis on multiple reactive environment, it was shown that the reaction strongly benefits from an early contact between guaiacol and the hydronium ion. This can occur more easily in a zeolite than in liquid water, leading to enhanced demethylation kinetics.

\section{Conclusions}
In this work, we presented ThermoLIB, a flexible Python/Cython library for the construction and post-processing of free energy profiles and surfaces from molecular simulation data. By formulating WHAM within a maximum-likelihood framework, ThermoLIB provides an analytical and broadly applicable estimation of uncertainties through the Fisher information matrix, yielding not only error bars but also full covariance information across the free energy surface. This enables rigorous error propagation toward derived thermodynamic and kinetic observables.

Beyond free energy reconstruction, ThermoLIB offers a unified framework to transform, project, and deproject free energy surfaces between different sets of collective variables, allowing users to diagnose sampling deficiencies, identify hidden variables, and iteratively refine their enhanced-sampling strategies. In addition, the library implements a rigorous transition-state-theory-based approach for extracting CV-independent kinetic rate constants, including consistent uncertainty estimates.

The tutorials and case studies demonstrate how ThermoLIB can be used to identify incomplete sampling, recover missing mechanistic information, and obtain reliable thermodynamic and kinetic quantities in complex chemical systems. By combining methodological rigor with practical usability, ThermoLIB aims to facilitate reproducible, statistically sound free energy analyses and to serve as a versatile post-processing platform for enhanced sampling simulations across a broad range of molecular applications.

\section*{Acknowledgements}
\footnotesize
M.B. wishes to thank the Research Foundation—Flanders (FWO) for a junior postdoctoral fellowship (grant n. 1269725N). The resources and services used in this work were provided by the VSC (Flemish Supercomputer Center), funded by the Research Foundation - Flanders (FWO) and the Flemish Government. L.V. acknowledges the Research Board of Ghent University (BOF). Furthermore, S. Borgmans and E. Van den Broeck are acknowledged for their help in testing and improving the library.

\section*{Supporting Information}
\footnotesize

\clearpage
\normalsize
\bibliography{bibliography}

\end{document}